# Synthesis of Ultra-Thin Superionic $Cu_2Se$ and New Aspects of the Low-Temperature Crystal Configurations

Abdulsalam Aji Suleiman[1*], Amir Parsi[1*], Mohammadali Razeghi[1*], Uğur Başçı[2], Saeyoung Oh[3], Doruk Pehlivanoğlu[2], Hu Young Jeoung[4], Kibum Kang[3], T. Serkan Kasırga[1,2α]

[1] Bilkent University UNAM – Institute of Materials Science and Nanotechnology, Ankara, 06800 Türkiye

[2] Department of Physics, Bilkent University, Ankara 06800 Türkiye

[3] Department of Materials Science and Engineering, Korea Advanced Institute of Science and Technology (KAIST), Daejeon 34141, Republic of Korea

[4] Graduate School of Semiconductor Materials and Devices Engineering, Ulsan National Institute of Science and Technology (UNIST), Ulsan 44919, Republic of Korea

[α] corresponding author e-mail: kasirga@unam.bilkent.edu.tr

*Equal contribution

**Abstract**

Superionic conductors offer unique advantages for novel technological devices in various fields, such as energy storage and neuromorphic computing. Above 414 K, $Cu_2Se$ turns into a well-known superionic conductor via a phase transition, and it is demonstrated to exhibit peculiar electrical and thermoelectric properties in bulk. Here, we report a large-area synthesis of ultra-thin single crystalline $Cu_2Se$ using the chemical vapor deposition method. We demonstrate that $Cu_2Se$ crystals exhibit optically and electrically controllable robust phase reconfiguration below 414 K. Moreover, our results show that the mobility of the liquid-like Cu ion vacancies in $Cu_2Se$ causes macroscopic fluctuations in the Cu ordering. Consequently, phase variations are not dictated by the diffusive motion of the ions but by the local energy minima formed due to the interplay between the extrinsic and the intrinsic material parameters. As a result, long-range ordering of the crystal below 414 K is optically observable at a micrometer scale. Our results show that $Cu_2Se$ could find applications beyond thermoelectric such as smart optical coatings, optoelectronic switching, and ionic transistors.

**Introduction**

Ionic motion within a solid is an intriguing phenomenon with many realized and potential technological applications. So far, many solids that exhibit superionic conductivity have been discovered that fall mainly into three different types, classified based on the nature of the phase transition from low to high conductivity state[1]. Among these ionic solids, cuprous selenide, $Cu_2Se$, has been an appealing material as it exhibits record high thermoelectric properties with a thermoelectric figure of merit, *zT*, reaching up to[2] 1.5 at 1000 K. Such high *zT* values are attributed to the high electrical transport with suppressed phonon transmission due to the disorder induced by the mobile Cu ions[3–7]. At room temperature, stoichiometric cuprous selenide is in the so-called β-phase where immobile Se ions form a superstructure of face-centered cubic (fcc) sublattice with the remaining Cu ions residing at multiple tetrahedral sites by forming various ordered configurations[8,9]. The crystal structure of the β-phase is considered to be monoclinic[10–13]. There are various face-centered-cubic-like structures with almost random Cu occupancy at various sites[8,14,15]. The difference between the total energy of various room temperature Cu configurations is determined to be very small by *ab initio* studies[11,15].

Above 414 K, the β-phase transforms to the α-phase through a continuous (second-order) phase transition[13,16–19]. Despite the confusion about labeling the low-temperature phase, as some papers call[20] it the "α-phase" we will stick with the conventional labelling[13,18,21–23]. Upon phase transition, Cu ions mobilize within the sublattice with the highest occupancy at the $8c$ tetrahedral (*T*) sites[18,22,23], along with a small fraction of the Cu ions can occupy the interstitial $4b$ octahedral (*O*) sites[24–27]. Also, $32f$ trigonal sites located around the *T* sites contribute to ion migration. Based on the X-ray diffraction measurements and *ab initio* molecular-dynamics simulations, it has been demonstrated that in cuprous selenide, Cu transport is possible via *T*→*O* hopping followed by an *O*→*T* hopping, whcih is energetically more favorable than *T*→*T* hopping[27]. As a result, fast ionic conduction is possible even in the stoichiometric $Cu_2Se$. **Figure 1a** depicts the structure of the β-phase and the positions of the *T* and *O* sites. Here trigonal sites are not shown to avoid confusion, but four interstitial positions exist per each *T* site. The phase transition, combined with the highly mobile Cu ions within the crystal, results in exceedingly complicated temperature-dependent structural phase dynamics that are not fully understood so far[21].

Despite the progress, all these studies have been performed on bulk single crystals, nano-powders, or polycrystalline samples of the cuprous selenide. The microscopic ordering in $Cu_2Se$ is evident in recent high-resolution transmission electron microscopy studies[8,28,29], and the microscopic domain configuration requires uniform samples for reproducible characterization of their properties. Even the most fundamental aspects regarding the crystal structure, phase transitions, as well as the nature of the very high thermoelectric figure of merit $zT$ are under debate[22,23,27,30–32]. It is well-established in other phase-change materials that samples larger than the characteristic domain size result in averaging of the observed effects and irreproducibility due to the sensitivity to the microscopic details of the samples to the intrinsic and extrinsic factors[33–35]. Thus, having single crystalline $Cu_2Se$ samples is imperative to avoid irreproducibility across the samples due to the broadening and hysteresis of the characteristics.

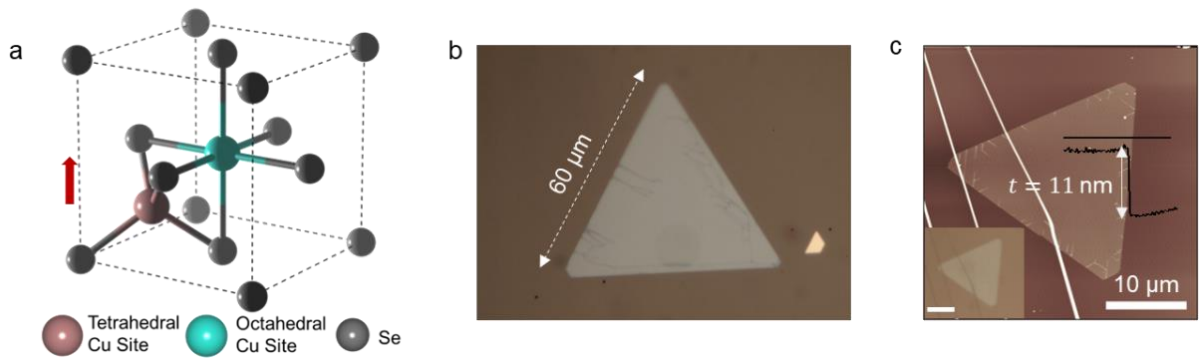

**Figure 1 a.** Unit cell of the room temperature $Cu_2Se$ with Cu, Tetrahedral, and Cu, Octahedral sites depicted. The red arrow points along the monoclinic [001] direction. **b.** Optical micrograph of a large $Cu_2Se$ crystal on a sapphire substrate. **c.** AFM height trace map of an 11 nm thick $Cu_2Se$ with the line trace overlayed on the image. The bright lines going across the image are the folds of the mica substrate. Inset shows the optical micrograph. The scale bar is 10 μm.

In this work, we synthesized ultra-thin, large-area $Cu_2Se$ single crystalline sheets down to 11 nm in thickness using the real-time optical chemical vapor deposition method[36] to study the exciting superionic properties while avoiding the aforementioned problems associated with the interfaces introduced due to sample polycrystallinity. Here, we focused on the growth mechanism of the crystals and characterized the structural, optical, and electrical properties of the $Cu_2Se$ crystals. We experimentally determined that the low-temperature stoichiometric β-phase exhibits three different configurations with different optical contrasts, which we labeled as "dark", "high-T dark", and "bright", and these states can be transformed reversibly and continuously to each other either via Joule or laser-induced heating. We find that the dark state forms due to the Cu vacancy formation at the edges of the crystals as a result of selective oxidation. At high temperatures, vacancies are distributed through the stoichiometric bright state and form the high-T dark state. Our electrical measurements show the metallic nature of the $Cu_2Se$. Inelastic laser spectroscopy measurements demonstrate that the dark state exhibits strong hot photoluminescence despite its metallic nature.

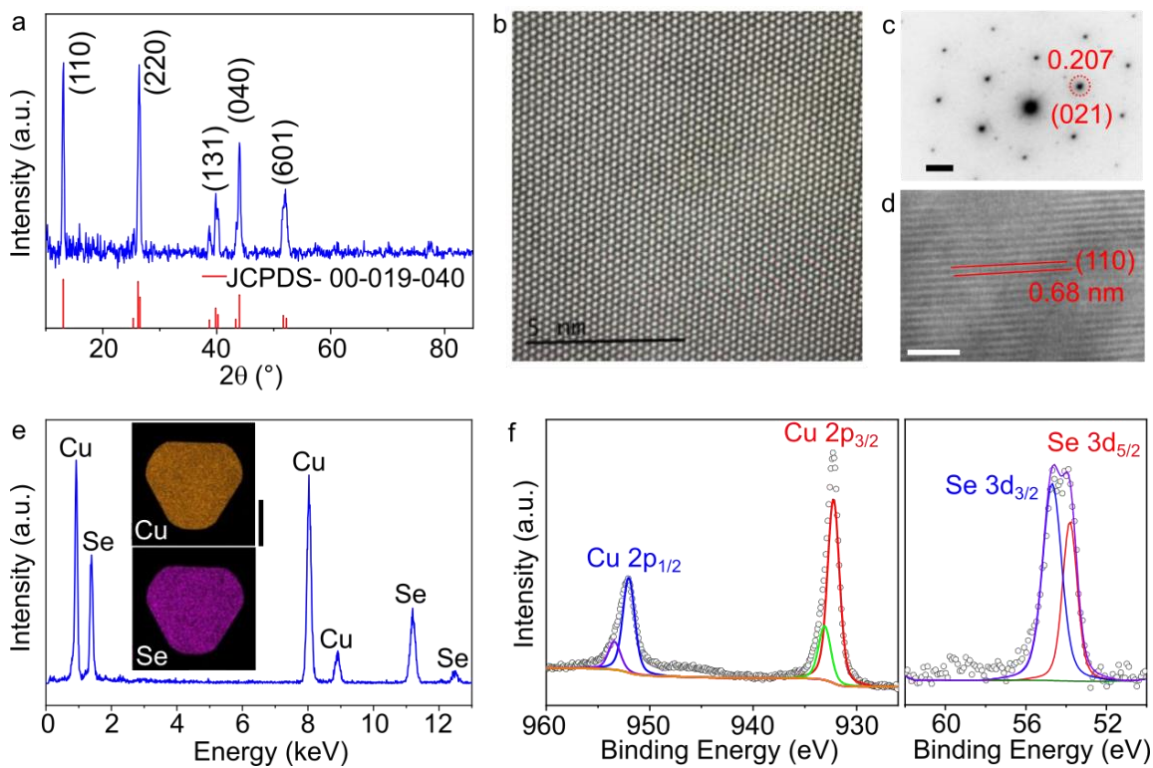

**Figure 2 a.** XRD θ-2θ scan of the $Cu_2Se$ crystals at room temperature. Red lines indicate the peak positions in JCPDS-00-019-0401 data card. **b.** HAADF STEM image of the crystal collected perpendicular to the crystal. **c.** SAED pattern indexed along [111] zone axis. The scale bar is 2 1/nm. **d.** Cross-sectional high-resolution TEM image of the crystal shows the interplanar spacing along (110) planes. The scale bar is 4 nm. **e**. EDX survey of a $Cu_2Se$ crystal with EDX maps overlayed on the graph. The scale bar is 2 μm. **f**. XPS surveys of Cu 2p and Se 3d peaks. The smaller peaks fitted to the Cu survey correspond to $Cu^{2+}$ oxidation state.

**$Cu_2Se$ Synthesis and Characterization**

Ultra-thin, large-area $Cu_2Se$ single crystal synthesis recipes are developed using the real-time optical chemical vapor deposition method (RTO-CVD) in which the synthesis of the crystals can be observed optically in real-time[36–38]. Se and CuCl in powder form are used as the precursors and Ar as the carrier gas. Typically, the growth takes place at 700 °C followed by a natural cooling down to room temperature. The synthesis can be performed on c-cut sapphire or fluorophlogopite mica substrates with a higher yield of thin crystals in the latter. We have also synthesized some thick crystals with very low yields on oxidized Si chips. Most of the crystals reported in this work are synthesized in a conventional CVD setup with a split-tube furnace by using the recipes from the RTO-CVD. **Figure 1b** shows the optical microscope image of a typical crystal and **Figure 1c** shows the atomic force microscope height trace map of an 11 nm thick $Cu_2Se$ crystal.

To understand the crystal structure and stoichiometry of the crystals, we performed grazing incidence X-ray diffraction (GI-XRD) on the samples. **Figure 2a** shows the θ-2θ scan plot. The peaks match very well with the orthorhombic crystal system of $Cu_2Se$ as reported in the literature[13,18] and referenced from the data card JCPDS-00-019-0401. The presence of multiple peaks corresponding to different crystal planes might be due to the presence of crystals grown on different crystallographic orientations on the substrate. This is not unexpected as $Cu_2Se$ has a highly symmetric structure and the growth kinetics are dictated by the wettability of the substrate surface by the reactant precursors rather than the crystal formation energy difference along various crystallographic directions. Moreover, the presence of various ordering of the Cu ions at the room temperature structure may also lead to the observed XRD pattern as shown in Figure 2. of Ref. [15].

High-angle annular dark field scanning TEM (HAADF STEM) and selected area electron diffraction (SAED )from samples transferred to holey carbon grids show regular atomic arrangement that agrees well with the XRD measurements (**Figure 2b**). Supercell reflections are also evident in the SAED given in **Figure 2c**. This hints at a $Cu^+$ vacancy ordering as reported before[39,40]. These vacancies form an ordered super-structure with a periodicity of 0.68 nm along the [111] crystallographic axis. Cross-sectional transmission electron microscopy (TEM) given in **Figure 2d** shows the interplanar spacing of 0.68 nm for the (110) plane. This matches well with the XRD measurements. EDX survey given in **Figure 2e** shows the presence of Cu and Se, exclusively. EDX maps are also provided as an inset in the figure. X-ray photoelectron spectroscopy (XPS) reveals that the Cu ions are in $Cu^+$ oxidation state predominantly with a slight contribution from $Cu^{+2}$ (**Figure 2f**). Ar ion milling for a few seconds removes the $Cu^{+2}$ peaks, showing that the crystals are susceptible to oxidation, consistent with the TEM measurements. Se 3d peaks can be fitted with two Gaussians corresponding to the Se $3d_{3/2}$ and $3d_{5/2}$ peaks.

**Low-temperature phase configurations of $Cu_2Se$**

Now, we will focus on the low-temperature configurations, i.e., below 414 K, of $Cu_2Se$. We observe three different optically contrasting regions which we label as the dark, high-T dark and bright. **Figure 3a** and **Figure 3b** show room temperature optical micrographs of two $Cu_2Se$ crystals grown on mica and sapphire substrates with optically contrasting regions agglomerated at the corners (mica) and spread irregularly (sapphire) over the crystal, respectively. We shall focus on samples on mica as it is easier to track the changes in the contrasting region configuration. We hypothesize that the bright state is the stoichiometric $Cu_2Se$ whereas the dark state is a macroscopic ordering of different Cu configurations at room temperature due to the Cu vacancy as a result of edge oxidation. Consequently, the high-T dark state is the vacancy-diffused bright state.

The room temperature polymorphism has been predicted earlier by *ab initio* studies[8,11,15] and the structures are derived from the β-phase by modifying the (001) plane of the pseudo-fcc sublattice of Se atoms in various configurations. Unlike the high-temperature phase, Cu ions in the β-phase occupy the *T*-sites in a variety of ordered configurations. Despite the small energy difference between the ordered configurations, long-range ordering-induced aggregation is possible[9] and has been observed in ultra-small clusters[40] of $Cu_2Se$ and hydrogen-doped $VO_2$ nanobeams[41–43].

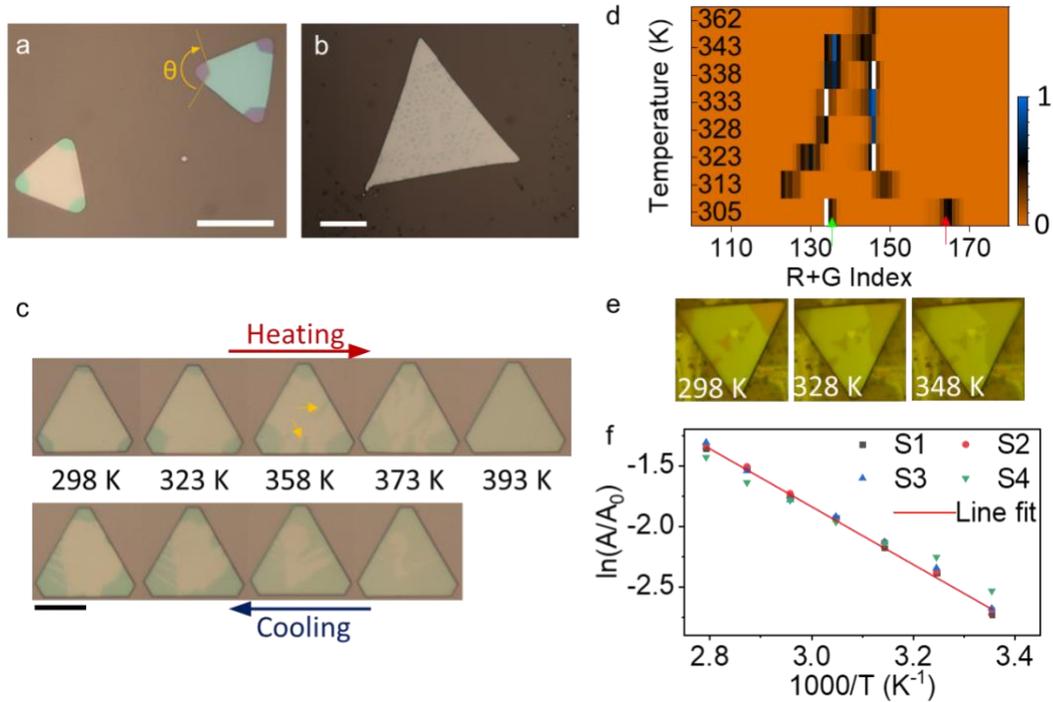

**Figure 3 a.** Optical micrograph of $Cu_2Se$ on a mica and **b.** on a sapphire substrate at room temperature. The optically contrasting regions are evident in the images. The sharp angle indicated on the crystal is 130° and measures the same for other crystals. Scale bars are 20 and 10 μm, respectively. **c.** A series of temperature-dependent optical micrographs of a crystal on mica. Red and blue arrows indicate the heating and cooling with the temperatures marked between the panels. There is a hysteresis between the area covered by the dark phases upon heating and cooling cycles. The yellow arrows indicate the filament-like dark phase formation at 358 K. The scale bar is 5 μm. **d.** 8-bit R+G index read from the dark regions of the sample at different temperatures. Green and red arrows mark the green and red indices, respectively. As the temperature increases green and red indices shift. This shows that there are two optically distinguishable dark configurations. **e.** Optical microscope images of a crystal taken at various temperatures show the shift of the color tint. The edge of the crystal is 19 μm long. **f.** Arrhenius plot of the dark phase area over the total area of the crystal versus $T^{-1}$ for four different samples of similar size. The activation energy is calculated from the slope of the line fit to be $E_A = 0.207 \pm 0.053$ eV.

To gain a deeper understanding of the nature of the bright and dark configurations, we observed the change of the optical contrast with the temperature, shown in **Figure 3c,** as a series of pictures of a typical crystal taken while heating and cooling. As the temperature increases, the dark state formation from the corners expands while its color tint shifts. Upon cooling to room temperature, the dark area is significantly larger than the initial condition. This is consistent with further oxidation of the samples through the edges at elevated temperatures as the measurement is performed under ambient. When we repeat the temperature cycling under vacuum, we find the dark configuration area to be the same as the initial condition (see **Figure S1** and **Figure S2**).

We also would like to note that there is a hysteresis between the heating and cooling cycles. Upon cooling, the bright configuration emerges below ~360 K.

To quantify the color tint shift, we averaged the red and green values of the 8-bit color index from a uniform region of the dark configuration on blue-filtered optical images. The red+green index versus temperature is given in **Figure 3d**. A clear shift in color values of the dark region is visible as the temperature increases. After the first jump in values after 305 K, the green index shifts towards higher values, and above 343 K, it merges with the red index. The change in the color tint clearly shows the presence of a high-T dark configuration. **Figure 3e** shows a series of optical microscope images of a crystal at various temperatures. The color tint of the shift of the dark configuration is very clear. A high-pass filter with a cut-off at 532 nm is used to enhance the contrast between the configurations.

We measured the temperature ($T$) dependence of the ratio of the area of the dark region ($A$) to the total area of the crystal ($A_0$). **Figure 3f** shows the Arrhenius behavior of $A/A_0 \propto e^{-\frac{E_A}{k_B T}}$ where $E_A$ is the activation energy and $k_B$ is the Boltzmann constant, with $E_A = 0.207 \pm 0.053$ eV. This activated behavior is consistent with the thermally activated hopping of copper vacancies between different positions of the tetrahedral sites. Moreover, the measured activation energy is very close to the total energy difference between polymorphs and the β-phase[11,15]. These observations and the measurement of the activation energy of the various configurations are consistent with the long-range ordering of the β-phase configurations predicted earlier. Interestingly, we find that the domain size of the order is almost as comparable to the size of the crystal.

**Inelastic spectroscopy on low-temperature configurations of Cu$_2$Se**

We performed Raman and photoluminescence spectroscopy on our samples at room temperature as well as at elevated temperatures up to 454 K. A micro-Raman/PL setup equipped with a 532 nm laser beam that can be focused to a spot at the diffraction limit is used for the measurements. First, we compared the spectra of the dark and bright regions at room temperature collected with 10 µW excitation power. As discussed later, we kept the laser power low to avoid the formation of the dark or the high-T dark phase. The dark region, as marked in **Figure 4a,** shows a broad photoluminescence centered around 2.1 eV (~600 nm), while the bright region shows a single Raman peak around 205 cm$^{-1}$ (~538 nm in the graph) over an elevated featureless background. We attribute the enhanced PL signal to the hot photoluminescence[14,44], which is prominent in metals, as discussed in the following paragraphs. PL mapping also matches well with the dark configuration distribution over the sample.

Next, we performed temperature-dependent PL measurements from room temperature to 454 K. **Figure 4b** shows the PL spectra collected both from the bright and the dark configurations. PL intensity from the dark region decreases while it remains the same for the bright region with the increasing temperature. We also notice that two different peaks can be fitted to the PL spectrum at low temperatures. At 348 K, 10 µW laser power is sufficient to manipulate the dark configuration in the bright regions of the crystal. In conjunction with the PL experiments, we studied the optical manipulation of crystal configurations.

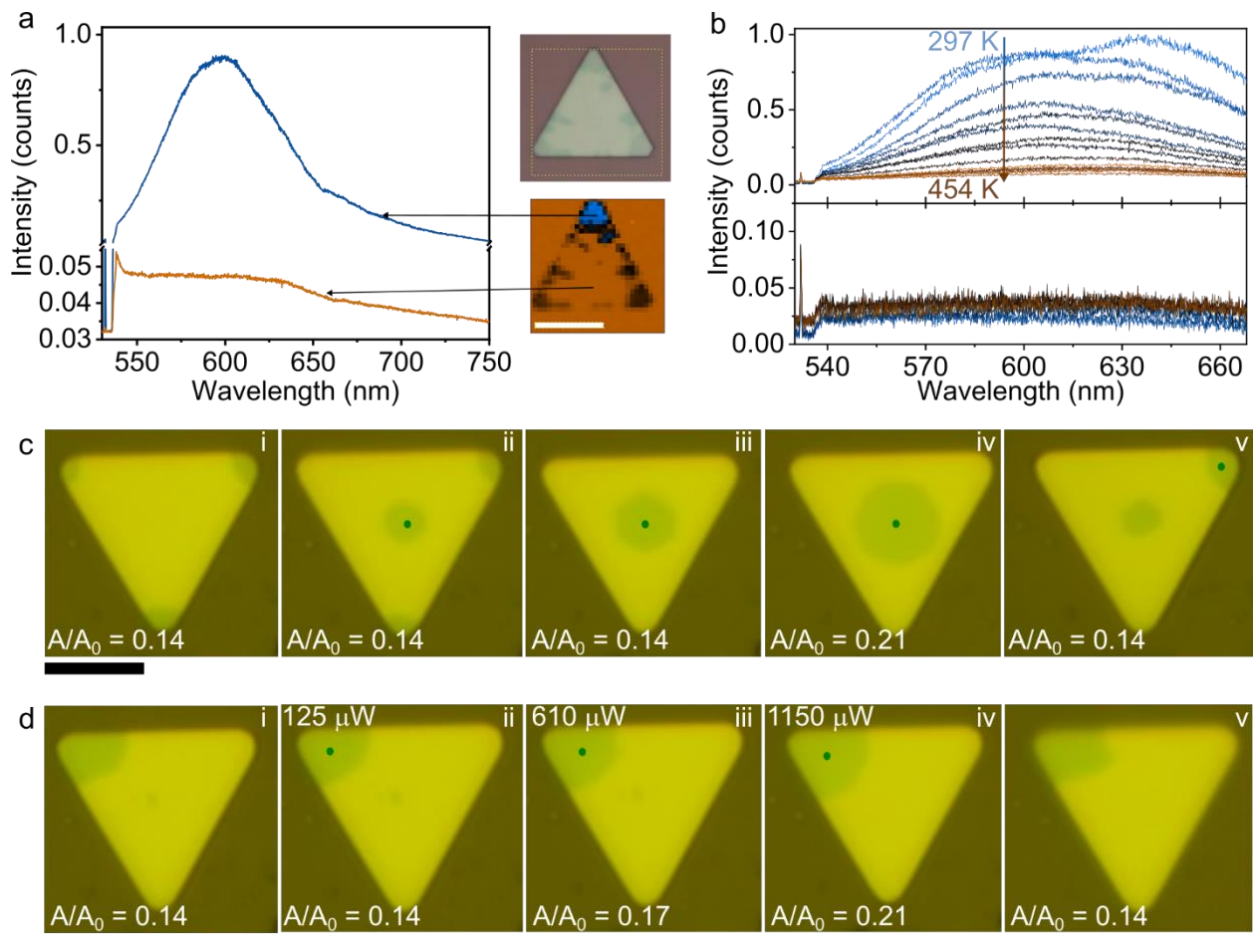

**Figure 4 a.** Photoluminescence spectra of the dark and the bright configurations are shown. The panel on the right shows the optical microscope image and the PL intensity map around 600 nm of the dashed region. The scale bar is 10 μm. **b.** Temperature-dependent PL spectra of the dark (upper graph) and the bright (lower graph) are given. Intensity is normalized to 1. **c.** Optical microscope images of **i.** pristine crystal with dark configurations at the corners, **ii.** the laser beam marked with the green dot focused on the center of the crystal, attracts the dark configuration to the center, **iii.** the dark configuration is completely pulled at the center. **iv.** Laser power is increased to ~1 mW. The dark area expands to a point where further increase of the laser power doesn't change the $A/A_0$ ratio. **v.** Once the laser power is reduced and the laser is positioned at the corner of the crystal, the dark region is moved to the corner, partially. The scale bar is 5 μm. **d.** The same crystal in **c** is shown. The dark phase is collected at the left upper corner of the crystal. The series of images from **i** to **iv** shows the effect of laser power increased when the laser is over the dark phase. **i.** From 0 μW to 125 μW, we observed no change in $A/A_0$ ratio. **ii-iv.** $A/A_0$ ratio increases with the increasing laser power. Beyond 1150 μW, no change in $A/A_0$ ratio is observed. **v.** Once the laser beam is blocked, the $A/A_0$ ratio shrinks to the pristine value.

**Optical manipulation of the crystal configurations**

As mentioned in the previous paragraph, the dark configuration can be manipulated using a focused laser beam at very low laser power. We performed a series of laser control experiments at room temperature to develop a better understanding of the optically contrasting regions. **Figure 4c** and **4d** shows a series of optical microscope micrographs of a crystal manipulated with a focused laser beam, taken under the micro-PL microscope with the 532 nm high pass filter inserted. **Figure 4c-i** shows the crystal before laser manipulation. In **Figure 4c-ii** and -**iii**, the laser is focused on the center of the crystal and its power is gradually increased to collect the dark configuration to the center. The dark phase configuration is changed even with a relatively low laser power (125 µW) and remains in the new configuration even after the laser is off. The fluid-like motion of the phase configuration shows the long-range ordering of the low-temperature configurations (see Supporting Video). One peculiar aspect of the laser-induced reconfiguration of the dark phase is that the ratio of the dark area to the total area ($A/A_0$) of the crystal remains unchanged. The ratio for the crystal in **Figure 4c-i** to -**iii** remains at 0.14 regardless of the reconfiguration. We observed this invariance at different temperatures as well (**Figure S2**). Once the laser power is increased even further (~1 mW) the dark configuration expands to $A/A_0 = 0.21$ (**Figure 4c-iv**). Beyond 1 mW, the ratio remains constant and when the laser is turned off, dark configuration area shrinks to its original size (**Figure 4c-v**). Only after applying larger laser powers (above 100 µW) for tens of minutes to several hours, depending on the laser power on the sample, the dark area expands irreversibly.

Next, **Figure 4d** shows a series of optical micrographs of a crystal manipulated with a laser. First, we collected all the dark contrasting regions to a corner using 100 µW total power on the sample. Then, by gradually increasing the laser power, we measured the total area of the dark region. Interestingly, up to 125 µW, no change in the dark area size is observed. From 125 to 1150 µW, the dark area increased while exhibiting a slight shift in the tint for the studied sample. Any increase in laser power beyond 1150 µW did not cause an increase in the dark configuration area as in **Figure 4c-iv**. When the laser beam is blocked, the dark area is reverted to its original size. These observations show that the dark contrast area is not determined by the temperature distribution created by the laser beam but rather by the local configuration altered by laser-induced heating, which mediates the reconfiguration of the vacancy-driven crystal configuration in a long range. Thus, further increases in the laser power cause no change in the dark region configuration.

**New aspects of the crystal configurations in Cu$_2$Se**

The measurements of PL and laser manipulation raise the following questions to be answered for a better understanding of the low-temperature configurations in Cu$_2$Se: (1) Why is the dark configuration located at the corners of the crystal in pristine samples at room temperature? (2) Why does $A/A_0$ ratio not change even though the dark contrast distribution changes after laser manipulation? (3) Why does $A/A_0$ ratio increase with the temperature?

To begin answering the questions, we make the following ansatz. After the growth of the crystal, a small amount of copper at the edges of the crystal leaves the lattice by oxidation after exposing the sample to the atmosphere. This creates a copper vacancy in the lattice, and its effect is most dominant at the corners, as the corners of the crystal have the largest surface-to-volume ratio as compared to the other parts of the crystal. Also, the edges of the Cu$_2$Se have a wedge shape facet formation (see **Figure S3**), and different crystallographic orientations may lead to different oxidation rates as reported for other materials[45]. Faster oxidation through the edge surfaces combined with the larger surface-to-volume ratio may lead to a larger copper deficiency at the corners as compared to the edges and the center of the crystal. The copper vacancy ordering stabilizes different Cu configurations at lower temperatures. This ansatz is consistent with our XPS measurements as well as previous reports in the literature[46] and explains question (1). Also, samples removed from the inert atmosphere of the growth chamber at room temperature show no dark contrast at the corners, while samples removed around before the complete cooldown of the furnace (~350-450 K) show larger dark contrast regions at the corners. However, a careful microscopic study is required to elucidate the underlying mechanism, and the extent of the copper vacancies should be further investigated.

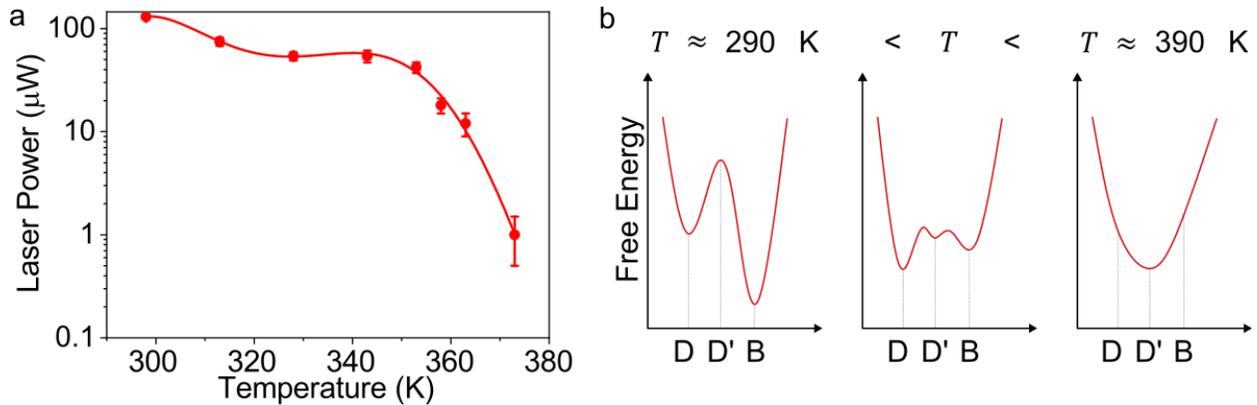

**Figure 5 a.** Minimum laser power required to create the dark and the high-T dark configurations from the bright configuration. The solid line shows a polynomial fit to aid the eye. **b.** Free energy configurations at three different temperature regimes are given qualitatively in the graphs. D, D', and B correspond to the dark, high-T dark, and the bright configurations, respectively. Near room temperature, by introducing Cu vacancies through oxidation, the D configuration can be obtained. As the temperature increases, D' configuration becomes meta-stable and can be obtained by locally heating the D or B configuration. Above 390 K, only D' configuration remains stable.

To answer the question (2), we would like to discuss how dark configuration emerges by laser illumination. First of all, there is a temperature-dependent laser power threshold to induce dark configuration. **Figure 5a** shows the change of the threshold laser power vs. temperature of the crystal. The decrease in the threshold laser power hints that the energy required to induce the dark configuration decreases with the temperature. When the laser beam heats a region, new Cu configurations are attained locally. However, as the rest of the crystal is still at ambient temperature, the pressure from the laser-heated region pushes the copper ions to the corners and converts the dark configuration to the bright configuration by enlarging the dark configuration in the vicinity of the laser spot. When the laser intensity is further increased, we observe a second enlargement of the dark region and the tint changes to the high-T dark configuration. Once this configuration is attained, the expansion stops. This clearly indicates that higher temperatures are required for the high-T dark configuration, and the finite boundary of the configuration shows that there is a limiting factor for the further increase of the boundary of the dark region.

The answer to question (3) includes hints from the answer to question (2). Even though the dark configuration at the corners of the sample is stabilized by the Cu vacancy, the dark configuration is more stable at elevated temperatures. Moreover, the copper vacancy diffusion rate increases with temperature. Thus, increasing the temperature results in the diffusion of copper vacancies into the crystal and stabilizes larger areas of the dark configuration. We can draw a qualitative free energy landscape for different Cu configurations, as in **Figure 5b**. The schematic shows how Cu vacancy alters the energy landscape and stabilizes various low-temperature crystal configurations.

**Electrical properties of Cu$_2$Se**

To gain a better understanding of the complex interplay of ionic motion with the crystal lattice in Cu$_2$Se, we fabricated two terminal devices of Cu$_2$Se. Device fabrication steps and device details are given in the methods section. When the devices are measured for the first time, typically, they exhibit very large resistance on the order of mega-Ohms with a non-Ohmic response (**Figure 6a**). When a bias above a certain threshold, typically around 1 V with a compliance of 1 mA, is applied, the resistance drops by almost 6 orders of magnitude (**Figure 6b**). A similar resistance drop is also observed on samples that are illuminated by a weak laser beam focused on the samples (**Figure S4**). We attribute the change to the drop in the contact resistance as a result of the ion migration-mediated conductive filament formation through the surface oxide layer, as the high resistance state is not retrievable in Au or In contacted devices. Temperature-dependent resistance measurements, given in **Figure 6c**, exhibit low resistivity and a positive temperature coefficient of resistance which are consistent with the metallic nature of the Cu$_2$Se, as reported earlier on Cu$_2$Se synthesized via other methods[18,47,48].

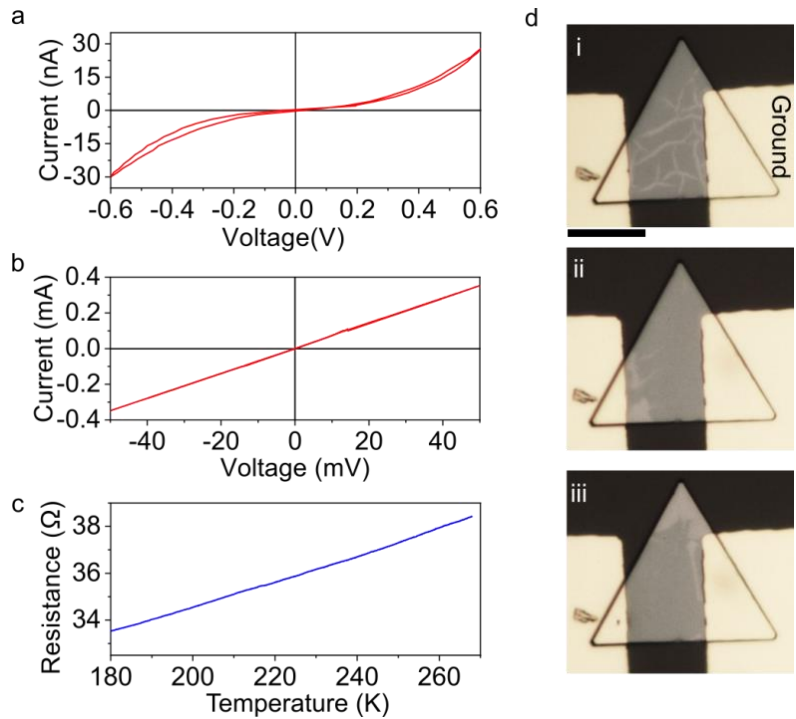

**Figure 6 a.** *I-V* curve before applying the forming voltage, 1 V with 1 mA compliance. **b.** *I-V* curve of the same device after forming the device. **c.** The resistance-Temperature graph shows a positive temperature coefficient of resistance. **d.** A series of optical micrographs were collected during the *I-V* cycling. Panel **i** shows the optical contrast of the crystal before the application of the forming voltage. **ii** shows after forming the device and **iii** when a reverse bias is applied. The ground terminal is marked on the first panel. The scale bar is 10 μm.

Optical changes in the phase configuration are simultaneously monitored during the *I-V* measurements. **Figure 6d** shows the optical micrographs of three major contrast changes observed optically in the device. **Figure 6d-i** shows the device before applying large voltages. The phase configuration is affected by the device fabrication steps. **Figure 6d-ii** shows the phase configuration after applying 1 V. **Figure 6d-iii** shows another phase contrast distribution after the application of reverse bias. The motion of the dark contrasting region is consistent with the motion of Cu vacancies as discussed in the previous section. Such mobilization of the different regions is consistent with the superionic nature of the $Cu_2Se$. The Supporting Video shows the real-time changes recorded under the optical microscope during *I-V* cycling.

**Conclusion**

In conclusion, we demonstrated the synthesis of large-area ultra-thin $Cu_2Se$ crystals on sapphire and mica substrates. Our results show that copper vacancies created by the edge-oxidation of the Cu in single-crystalline $Cu_2Se$, result in an optically distinguishable order that can be manipulated electrically and optically. Among our results, we show that the dark configuration has a strong broad photoluminescence that can be attributed to hot photoluminescence. The ability to control the vacancy order on the length scale of crystallite dimensions may enable novel applications in optical switching based on superionic transport. These results show $Cu_2Se$ can be considered for applications beyond the thermoelectrics[2,20,49,50].

**Methods**

Substrates (sapphire or $SiO_2$) were ultrasonically pre-cleaned in isopropanol and deionized water and dried using a high-purity $N_2$ spray gun and the mica is freshly cleaved with a surgical blade with no additional cleaning process. The target substrate is placed downstream of the tube and 10 mg Se powder is in a crucible upstream. Then CuCl powder (10 mg) is placed in a separate crucible placed in the center of the furnace about 1 cm away from the substrate. The system is pumped down to $2.0 \times 10^{-2}$ mbar, then purged with Ar gas for five minutes and heated to the optimum growth temperature (600 °C, 650 °C, and 700 °C for $SiO_2$, sapphire, and mica, respectively). During growth, ambient pressure was maintained in the system and the dwell time was set to 5 minutes. After growth, the system was cooled naturally to room temperature.

Two terminal devices are fabricated through standard optical lithography procedures followed by thermal evaporation of 5 nm Cr followed by 100 nm Au.

**Author Contributions**

A.A.S., A.P., and M.R. contributed equally to this work. A.A.S. and A.P. worked on the synthesis and structural characterization. M.R. performed the device fabrication as well as the temperature-dependent electrical measurements. T.S.K. conceived the experiments and wrote the manuscript. All the authors commented on the writing of the manuscript.

# Supporting Information for Synthesis of Ultra-Thin Superionic $Cu_2Se$ and New Aspects of the Low-Temperature Crystal Configurations

Abdulsalam Aji Suleiman[1*], Amir Parsi[1*], Mohammadali Razeghi[1*], Uğur Başçı[2], Saeyoung Oh[3], Doruk Pehlivanoğlu[2], Hu Young Jeoung[4], Kibum Kang[3], T. Serkan Kasırga[1,2α]

[1] Bilkent University UNAM – Institute of Materials Science and Nanotechnology, Ankara, 06800 Türkiye

[2] Department of Physics, Bilkent University, Ankara 06800 Türkiye

[3] Department of Materials Science and Engineering, Korea Advanced Institute of Science and Technology (KAIST), Daejeon 34141, Republic of Korea

[4] Graduate School of Semiconductor Materials and Devices Engineering, Ulsan National Institute of Science and Technology (UNIST), Ulsan 44919, Republic of Korea

[α] corresponding author e-mail: kasirga@unam.bilkent.edu.tr

*Equal contribution

**Temperature cycling of $Cu_2Se$ under vacuum**

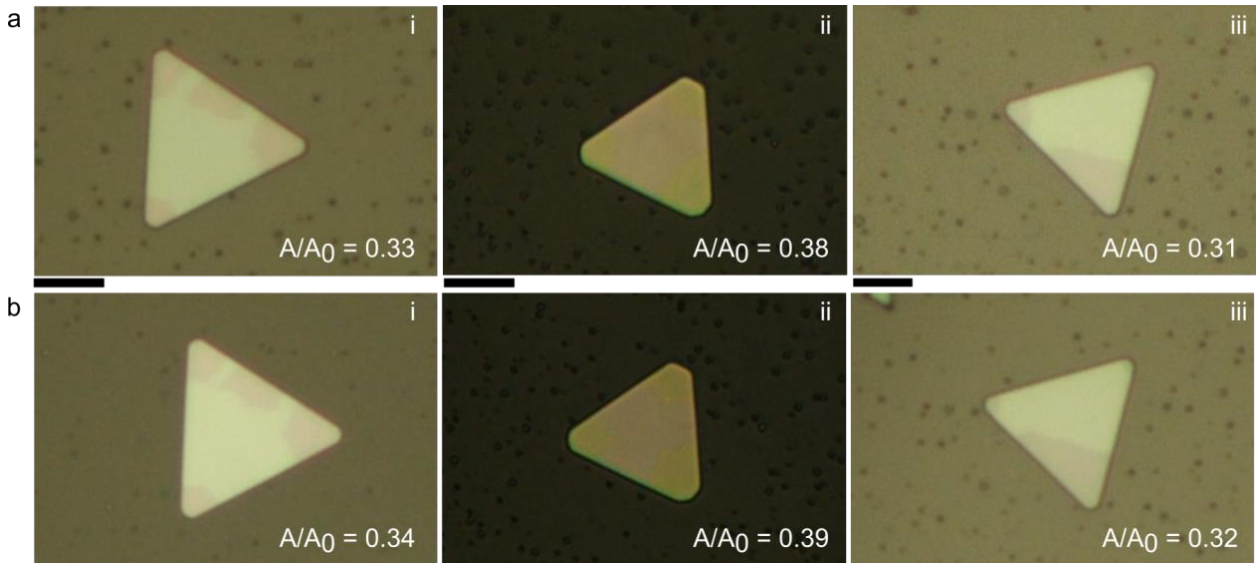

**Figure S1 a.** Optical micrographs of various $Cu_2Se$ crystals taken before temperature cycling to 400 K are given. **b.** Optical micrographs of the same crystals at room temperature, after heating to 400 K under $10^{-2}$ mbar vacuum. $A/A_0$ ratio is indicated in the figures. The scale bars for panels **i** and **ii** are 10 μm and for **iii** 5 μm.

**Temperature cycling under ambient and laser manipulation of the dark configuration**

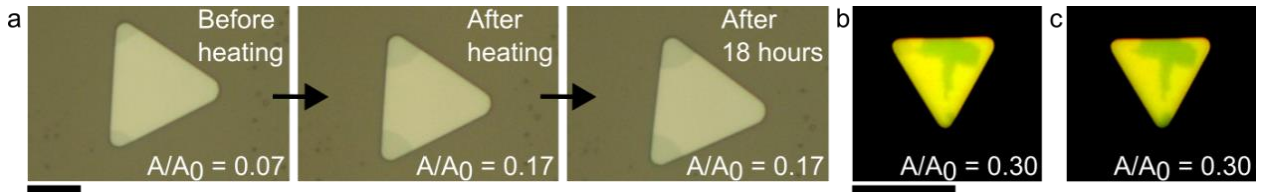

**Figure S2 a.** A series of optical micrographs of Cu₂Se crystal before heat cycling under ambient, after heat cycling above 400 K, and after keeping it at room temperature in ambient for 18 hours is shown. As discussed in the main text, $A/A_0$ ratio increases after teat cycling and remains unchanged at room temperature. **b.** Optical microscope micrograph of a crystal at 348 K before laser manipulation and **c.** after laser manipulation. $A/A_0$ ratio remains the same. The scale bars are 10 μm.

**SEM Images of Cu₂Se**

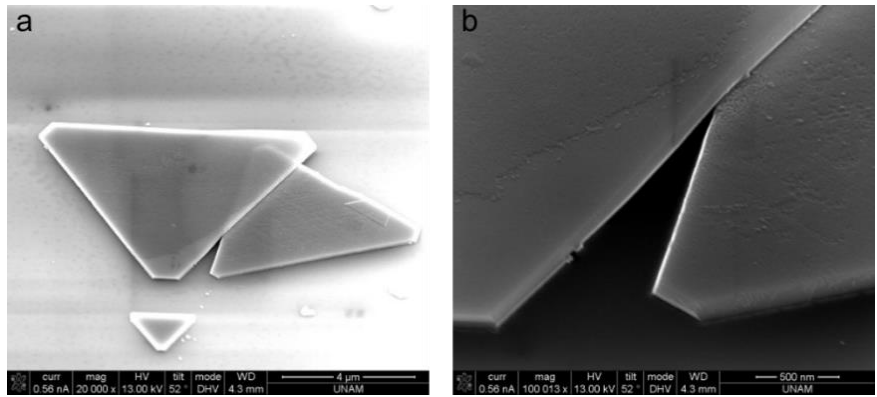

**Figure S3 a.** SEM image of Cu₂Se crystals transferred over SiO₂ substrate. Two crystals have landed on each other. **b.** A close-up SEM image of the edge of the crystals shows the wedge-shaped edges of Cu₂Se.

**Laser-induced change in the electrical conductivity of Cu₂Se**

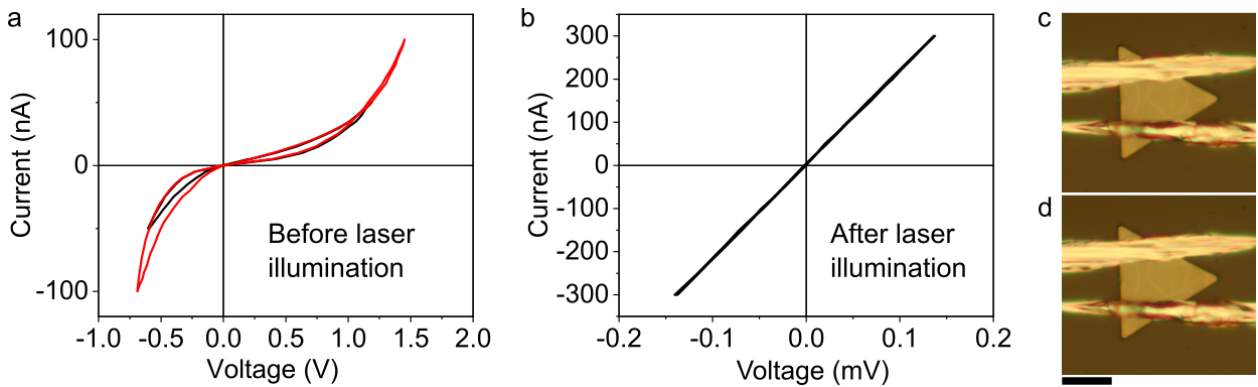

**Figure S4 a.** *I-V* graph of an indium-contacted Cu₂Se device before laser illumination and **b.** *I-V* graph of the same device after laser illumination. 212 μW laser power is used to partially illuminate the crystal. **c.** Optical micrograph of the device before illumination and **d.** after illumination is given. The scale bar is 10 μm.